\documentclass[conference]{IEEEtran}
\IEEEoverridecommandlockouts
\usepackage{cite}
\usepackage{amsmath,amssymb,amsfonts}
\usepackage{algorithmic}
\usepackage{graphicx}
\usepackage{textcomp}
\usepackage{xcolor}
\usepackage{svg}
\usepackage{tipa}
\usepackage{colortbl}
\usepackage{multirow,arydshln}
\usepackage{tablefootnote}
\usepackage{balance}
\usepackage{hyperref}
\def\BibTeX{{\rm B\kern-.05em{\sc i\kern-.025em b}\kern-.08em
    T\kern-.1667em\lower.7ex\hbox{E}\kern-.125emX}}
\begin{document}

\title{\fontsize{19}{24}\selectfont Addressing Emotion Bias in Music Emotion Recognition and Generation with Frechet Audio Distance}


\author{\IEEEauthorblockN{Yuanchao Li\thanks{Work done while Yuanchao and Azalea were interning at Microsoft Research. Correspondence to \textit{yuanchao.li@ed.ac.uk}.}}
\IEEEauthorblockA{\textit{University of Edinburgh, UK}}
\and
\IEEEauthorblockN{Azalea Gui}
\IEEEauthorblockA{\textit{University of Toronto, Canada}}
\and
\IEEEauthorblockN{Dimitra Emmanouilidou}
\IEEEauthorblockA{\textit{Microsoft Research, US}}
\and
\IEEEauthorblockN{Hannes Gamper}
\IEEEauthorblockA{\textit{Microsoft Research, US}}
}

\maketitle

\begin{abstract}
The complex nature of musical emotion introduces inherent bias in both recognition and generation, particularly when relying on a single audio encoder, emotion classifier, or evaluation metric. In this work, we conduct a study on Music Emotion Recognition (MER) and Emotional Music Generation (EMG), employing diverse audio encoders alongside Frechet Audio Distance (FAD), a reference-free evaluation metric. Our study begins with a benchmark evaluation of MER, highlighting the limitations of using a single audio encoder and the disparities observed across different measurements. We then propose assessing MER performance using FAD derived from multiple encoders to provide a more objective measure of musical emotion. Furthermore, we introduce an enhanced EMG approach designed to improve both the variability and prominence of generated musical emotion, thereby enhancing its realism. Additionally, we investigate the differences in realism between the emotions conveyed in real and synthetic music, comparing our EMG model against two baseline models. Experimental results underscore the issue of emotion bias in both MER and EMG and demonstrate the potential of using FAD and diverse audio encoders to evaluate musical emotion more objectively and effectively.
\end{abstract}


\section{Introduction}
\label{sec:intro}
Music Emotion Recognition (MER) and Emotional Music Generation (EMG) play crucial roles in various applications, such as music recommendation systems, film scoring, and emotion therapy. For instance, in music recommendation systems, accurate MER matches user preferences, increasing satisfaction. Similarly, in emotion therapy, generating music with emotion can enhance therapeutic effects.  

However, unlike speech emotion, which can be expressed through language (i.e., spoken content), the perception and expression of music emotion are more subjective and challenging, inevitably leading to biases in its measurement, recognition, and generation. In MER, the lack of alignment between categorical and dimensional models challenges the generalizability of current MER approaches \cite{hung2021emopia,yang2011music}. Furthermore, MER is sometimes referred to as \textit{music mood classification} when it employs five mood clusters as labels \cite{hu2007exploring}, which also introduces bias due to variations in human perception. Moreover, audio encoders are designed with distinct training objectives defined by humans, and relying on a single encoder for MER inevitably introduces inductive bias.

In EMG, there are two primary approaches to encoding emotion. The first approach uses emotion labels as conditioning, where the bias stems directly from the labels assigned by human annotators \cite{sulun2022symbolic}. In the second approach, which replaces the human annotator with an MER classifier, emotion bias is inductive and inherited from the classifier, largely influenced by the choice of audio encoders. Therefore, it is crucial to explore more objective approaches to measure, recognize, and generate music emotion, mitigating the aforementioned emotion bias. In this paper:

$\bullet$ We conduct a benchmark evaluation of MER for both categorical and dimensional emotions, utilizing various audio encoders and metrics to highlight the issue of emotion bias. We then introduce Frechet Audio Distance (FAD) as an objective measure to address this challenge.

$\bullet$ We propose an enhanced EMG approach to reduce emotion bias in emotion labels and improve the realism of generated emotional music. Using our approach, alongside two baseline methods, we generate three types of synthetic emotional music for comparison with real music. The effectiveness of our approach is demonstrated through evaluations using FAD scores calculated by multiple audio encoders.

\begin{figure*}[ht]
  \centering
  \includegraphics[width=\textwidth]{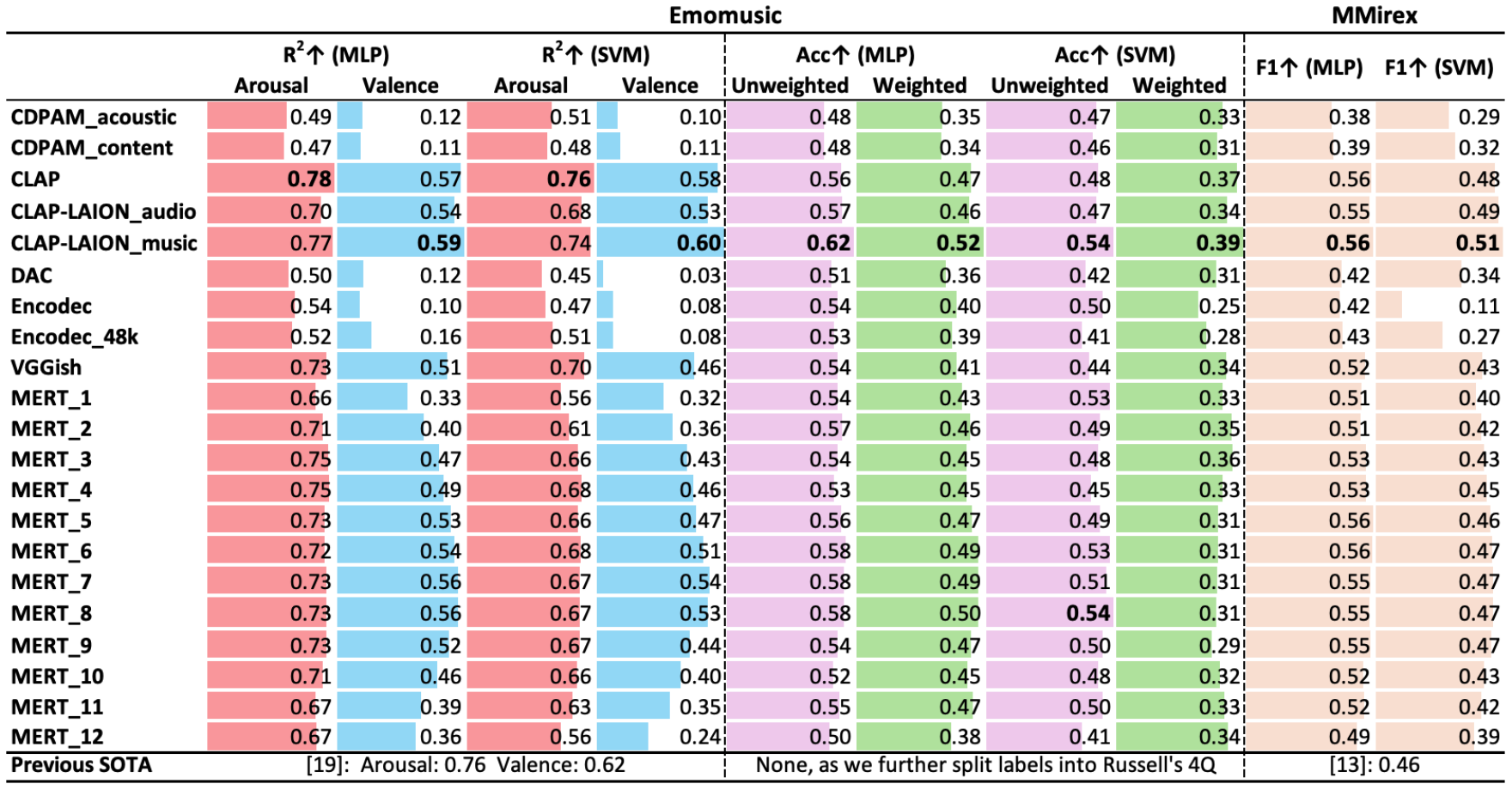}
  \caption{Comparison results of MER using different audio encoders on Emomusic and MMirex. $\uparrow$: the higher the better. Bold: best performance.}
  \label{fig:mer}
\end{figure*}

\section{Related Work}
\subsection{Emotion Bias in Music Emotion Recognition}
\label{sec:related}
Emotion bias in MER arises from various sources, including imbalances in training data, subjective emotion annotations, and cultural differences in emotional expression. Studies have shown that MER models trained on Western music often perform poorly when tested on non-Western music due to differences in emotional interpretation \cite{krumhansl2002music,yang2011music}. Similarly, relying on human-annotated labels introduces subjectivity, as listeners' emotional responses can vary significantly depending on their cultural background, personal experiences, and even their mood during listening sessions \cite{hallam2009oxford,hunter2010music}. These biases result in MER models that reflect the emotional norms of a specific population, thereby limiting the universality of the generated predictions. Recent efforts to address these biases include domain adaptation aimed at improving the transferability of MER models across different musical cultures \cite{chen2018cross}.

Moreover, while categorical and dimensional emotions are not entirely incompatible, their alignment is imperfect. For instance, Russell’s circumplex model (Russell's 4Q) \cite{russell1980circumplex} effectively captures specific emotions: joy exhibits positive valence and high arousal, while sadness displays negative valence and low arousal. However, it falls short in distinguishing between certain emotions. For example, both fear and anger share negative valence and high arousal. Such misalignment introduces emotion bias in measurement. For instance, an MER method may perform well for categorical emotions but struggle with dimensional emotions, posing challenges to the generalizability of existing MER approaches.

Additionally, bias can arise from the use of different audio encoders due to their distinct training strategies. For instance, features extracted by Encodec \cite{kang2023emogen} may lose subtle emotional nuances because of its audio compression training objectives. Moreover, there is no guarantee that an encoder optimized for categorical emotion classification will perform equally well for dimensional emotion modeling \cite{song2016perceived}.

\subsection{Emotion Bias in Emotional Music Generation}
EMG is similarly susceptible to emotion bias as MER. Since music generation relies on datasets for training, the emotional diversity of the generated music is inherently tied to the diversity of the training set. Models trained on datasets with emotional imbalances tend to generate music that predominantly expresses the more frequent emotions. Although efforts have been made to create more balanced datasets that cover a broader range of emotional contexts, this requires significant human effort.

Moreover, emotion bias can also arise during model training. As mentioned in Section~\ref{sec:intro}, there are two primary approaches to EMG. In the first approach, which relies on using emotion labels as control signals, subjectivity stems from the labels themselves, as they are assigned by human annotators \cite{sulun2022symbolic,madhok2018sentimozart}. In the second approach, which employs an emotion classifier for guidance, bias is inherited from the MER models due to the limitations of the audio encoders. Recent methods have introduced constraints into the generation process to ensure emotional variety and reduce bias in EMG. For instance, rather than directly encoding emotion embeddings from labels, \cite{kang2023emogen} mapped categorical labels into four quadrants, each corresponding to specific musical attributes (i.e., symbolic features) used for conditioning. However, this method relies on only four sets of attributes, with each quadrant representing a set of emotions. As a result, all emotions within the same quadrant share the same musical attributes as conditions, limiting the variation and naturalness of the generated emotions.

To address these issues, we aim to address emotion bias in both recognition and generation by \textit{1) measuring and recognizing music emotion in a more objective and efficient way} and \textit{2) generating music emotions that more closely resemble those in real music}.

\section{Datasets and Audio Encoders}
We adopt multiple datasets to ensure generalizability and for different purposes: two for recognition and one for generation.

\textbf{Emomusic} \cite{soleymani20131000} consists of 744 music excerpts, each 45 seconds long, annotated with two-dimensional Valence-Arousal (VA) labels. The dataset includes human singing across various genres, such as pop, rock, classical, and electronic.

\textbf{MMirex} \cite{panda2013multi} contains 903 audio samples, each labeled with emotion tags from the MIREX Mood Classification Task \cite{downie20082007}. The dataset uses five mood clusters as defined by MIREX and features audio from various instruments.

\textbf{EMOPIA} \cite{hung2021emopia} comprises 387 piano solo performances of popular music, manually segmented into 1,087 clips for emotion annotation. The emotion classes follow Russell’s 4Q model, which defines emotions within the VA space.

For audio encoders, we include models trained with various objectives, such as \textbf{VGGish} (convolutional embeddings) \cite{hershey2017cnn}, \textbf{CLAP} \cite{elizalde2024natural} and \textbf{CLAP-LAION} \cite{laionclap2023} (contrastive embeddings), \textbf{MERT} \cite{li2023mert} (self-supervised embeddings), \textbf{CDPAM} \cite{Manocha2021CCL} (audio similarity), as well as \textbf{EnCodec} \cite{defossez2022highfi} and \textbf{DAC} \cite{kumar2024high} (low-rate audio codecs). For models with two versions trained on different inputs or objectives (e.g., audio and music), we include both. For MERT, we compare all 12 hidden layers for layer-wise analysis (see Fig.~\ref{fig:mer}).

\section{Music Emotion Recognition}
In this experiment, we use the Emomusic and MMirex datasets. Based on the VA labels from Emomusic for dimensional MER, we further categorize them into Russell's four quadrants (4Q) for categorical MER: Q1 (-V+A), Q2 (-V-A), Q3 (+V+A), and Q4 (+V-A). Both dimensional and categorical MER are performed on this dataset. Additionally, MMirex is used for categorical MER based on five mood clusters. Thus, both dimensional (VA) and categorical (Russell's 4Q and five mood clusters) emotions are investigated to ensure generalizability.

For dimensional MER, we use the coefficient of determination ($R^2$) as the evaluation metric for VA. For categorical MER, we assess both weighted and unweighted accuracy on Emomusic, given its imbalanced distribution, and use the F1-score on MIREX, following the literature. We employ both SVM and Multi-Layer Perceptrons (MLP), while maintaining consistent experimental settings with previous research \cite{castellon2021codified}. The experiments are conducted using an NVIDIA A100 GPU, with models trained using the Adam optimizer at a learning rate of 0.0001 and a batch size of 32. All models are trained under the same conditions to ensure consistency. It can be observed from Fig.~\ref{fig:mer} that:

\textit{\textbf{1)}} Valence is more challenging to predict than arousal, which is consistent with previous research \cite{kim2010music}. While arousal is typically conveyed through prosodic features such as pitch, energy, and tempo, valence relies more heavily on the intricate aspects of musical composition, as noted by \cite{droit2013music}. Moreover, since speech valence is significantly influenced by language content \cite{li2019expressing}, this relationship suggests the existence of a distinct ``language'' within music, supporting the conclusions of \cite{mcpherson2018role}.

\textit{\textbf{2)}} For dimensional emotion, a significant performance gap is observed among audio encoders. However, this gap becomes less pronounced for categorical emotion, resulting in more comparable performance across different encoders. This phenomenon is evident in both categorical quadrants (EmoMusic) and categorical mood clusters (MIREX), illustrating the challenges posed by bias in MER, as discussed in Sec.~\ref{sec:intro}. Consequently, an encoder that performs well with categorical emotions does not necessarily excel at encoding dimensional emotions.

\textit{\textbf{3)}} Different encoders can exhibit significant performance differences. Those trained using contrastive audio-language learning methods, such as CLAP and CLAP-LAION, achieve the best performance on all metrics. However, CDPAM, which is also trained using contrastive learning but on the contrast between different audio samples rather than between audio and language, underperforms compared to most of the other encoders. This suggests that MER benefits from language-like information, indicating that music segments contain contextual meaning that can somewhat serve as a ``language'', which is consistent with the finding in 1). On the contrary, encoders designed for audio compression, such as DAC and Encodec, generally do not perform well in MER. This is likely due to audio compression leading to information loss related to emotion or the emotional information implicitly residing in the embeddings after compression.

\textit{\textbf{4)}} In the layer-wise results of MERT, we observe an upward-downward trend, consistent with previous studies on speech emotion recognition \cite{li2023exploration,saliba2024layer}, indicating the similarity between speech and music emotions. This trend becomes more pronounced when predicting valence, highlighting the need for deeper encoding to effectively capture valence. This observation further supports the finding that valence in music depends on higher-level musical structures, such as contextual elements.

Furthermore, the performance using $R^2$ and accuracy differs significantly, with unweighted accuracy notably outperforming weighted accuracy, indicating the impact of data balance. These findings reveal that emotion bias in MER can stem from various factors, including the audio encoder, evaluation metric, and emotion dimension.

\section{Measuring Music Emotion via FAD}
\label{sec:fad}
FAD was initially proposed for evaluating music enhancement quality and has shown a close correlation with human perception as a reference-free metric, without the need for model training \cite{kilgour2019frechet}. Recently, it has replaced humans in assessing acoustic and musical quality and similarity, helping to mitigate subjective bias. For instance, it has been shown to effectively distinguish between real and synthetic audio \cite{gui2023adapting}, as well as audio with different emotions \cite{sun2025exploring}. In light of this, we adopt FAD as an objective metric to resolve potential emotion bias arising from various sources (implemented using Microsoft FADTK\footnote{\href{https://github.com/microsoft/fadtk}{https://github.com/microsoft/fadtk}}).

FAD compares acoustic features generated from two audio sets, enabling it to assess two distinct audio sets without ground truth. The FAD score is calculated using multivariate Gaussians from two embedding sets $N_a(\mu_a, \Sigma_a)$ and $N_b(\mu_b, \Sigma_b)$ as follows:
\begin{align}
F(N_a, N_b) = ||\mu_a - \mu_b|| ^2 + tr(\Sigma_a + \Sigma_b - 2\sqrt{\Sigma_a\Sigma_b})
\end{align}
where $tr$ is the trace of a matrix. To calculate FAD scores, we extract features from the same audio encoders as used in Fig.~\ref{fig:mer} by computing the average of the FAD scores to mitigate the bias introduced by any single encoder. Since separating music sets based on continuous values is challenging, we only consider discrete labels. The results for Emomusic and MMirex are presented in Table~\ref{tab:fademomusic} and Table~\ref{tab:fadmmirex}, respectively.

\begin{table}[ht!]
\centering
\large
\caption{FAD scores on Emomusic (categorical quadrant).}
\scalebox{0.76}{
\begin{tabular}{lcccccc}
\hline
 & \textbf{Q1\_Q2} & \textbf{Q1\_Q3} & \textbf{Q1\_Q4} & \textbf{Q2\_Q3} & \textbf{Q2\_Q4} & \textbf{Q3\_Q4} \\ \hdashline
\textit{FAD} & \normalsize{8.69} & \normalsize{33.46} & \normalsize{16.64} & \normalsize{23.07} & \normalsize{33.88} & \normalsize{10.47} \\ \hline
\end{tabular}}
\label{tab:fademomusic}
\end{table}

From Table~\ref{tab:fademomusic}, we observe the following: \textbf{\textit{1)}} The scores are high between diagonal quadrants (i.e., Q1\_Q3 and Q2\_Q4). This is reasonable since the polarities of both valence and arousal are opposite in diagonal quadrants. \textbf{\textit{2)}} The score for Q1\_Q2 is lower than that for Q1\_Q4, and the score for Q3\_Q4 is lower than that for Q2\_Q3. This indicates that the difference between two music sets is smaller when they share the same arousal polarity but differ in valence polarity, compared to when they share the same valence polarity but differ in arousal polarity. These observations suggest that arousal is more explicitly expressed in music than valence.

\begin{table}[ht!]
\centering
\caption{FAD scores on MMirex (categorical cluster).}
\begin{tabular}{lccccc}
\hline
 & \textbf{C1\_C2} & \textbf{C1\_C3} & \textbf{C1\_C4} & \textbf{C1\_C5} & \textbf{C2\_C3} \\
\textit{FAD} & \normalsize{3.27} & \normalsize{9.08} & \normalsize{3.67} & \normalsize{4.34} & \normalsize{6.05} \\ \hdashline
 & \textbf{C2\_C4} & \textbf{C2\_C5} & \textbf{C3\_C4} & \textbf{C3\_C5} & \textbf{C4\_C5} \\
\textit{FAD} & \normalsize{3.28} & \normalsize{9.27} & \normalsize{5.24} & \normalsize{15.00} & \normalsize{7.61} \\ \hline
\end{tabular}
\label{tab:fadmmirex}
\end{table}

From Table~\ref{tab:fadmmirex}, we observe variations among mood clusters. The score for C1\_C2 is the smallest, while the score for C3\_C5 is the largest. Although mood clusters do not explicitly represent valence and arousal, we can infer that both C1 (e.g., \textit{Passionate}) and C2 (e.g., \textit{Cheerful}) are associated with high arousal, which explains the small FAD score between them. In contrast, C3 (e.g., \textit{Wistful}) appears to be linked with low arousal, while C5 (e.g., \textit{Aggressive}) is associated with high arousal, which justifies the relatively large FAD score. These findings are consistent with prior research on the relationship between categorical and dimensional music emotion \cite{hong2017analysis}. Due to space constraints, we omit detailed descriptions of the mood clusters and refer readers to \cite{hu2007exploring} for more information.

Thus, averaging FAD scores from various encoders offers a more objective manner for measuring emotions, eliminating the need to train MER models and select metrics, and thereby reducing bias. Furthermore, this approach addresses issues of imbalanced data.

\section{Emotional Music Generation}

After benchmarking emotion recognition and introducing the use of FAD for objective measurement, we extend this approach to the domain of music generation for emotion assessment. We propose an enhanced generation model that integrates both categorical and dimensional emotions to address their respective limitations, and then compare the music generated by our model with that produced by two baselines.

\subsection{Baseline Generation Models}
Since using emotion classifiers for guidance introduces additional emotion bias \cite{kang2023emogen}, we use emotion labels solely for conditioning. Two baseline models are selected for generating symbolic music: \textit{MIDIEmo} \cite{sulun2022symbolic} for dimensional emotion and \textit{EmoGen} \cite{kang2023emogen} for categorical emotion.

\textbf{\textit{MIDIEmo}} encodes continuous VA labels (e.g., [0.8, 0.8]) as latent embeddings using linear layers, which are then concatenated with music embeddings. We argue that this approach can make music samples with boundary labels more susceptible to subjective bias (as the boundary emotions are usually difficult to distinguish). The challenge of assigning boundary labels with high confidence often leads to ambiguity in the generated emotion. Consequently, while there is considerable variation, ambiguity is prevalent in the generated music, making it difficult to perceive precise emotions.

\textbf{\textit{EmoGen}} uses four pre-extracted embeddings to represent emotions within the four quadrants. It encodes discrete values of these quadrants (e.g., 1 or 2) into embeddings as the emotion input, thus samples within a particular quadrant share the same emotional conditions (i.e., the emotion embeddings for all samples in each quadrant are identical, regardless of their original labels). As a result, while the generated music exhibits precise emotions, we argue that it lacks variation, leading to a situation where all samples with the same emotion tend to sound emotionally similar.

\subsection{Proposed Generation Model}

Instead of using either continuous VA values or discrete emotion quadrants alone, we combine them in the conditioning process. First, we determine the quadrant based on the continuous VA values (e.g., +V+A corresponds to Q1). Next, we encode both the quadrant and the VA values into hidden embeddings using separate linear layers. Subsequently, we merge these embeddings via a weighted sum to create the final emotion embedding:
\begin{equation}
\begin{aligned}
Emotion &= wgt_{q} \cdot embd_{q} + (1-wgt_{q}) \cdot embd_{va}
\end{aligned}
\label{eq:weightedsum}
\end{equation}
where $embd_{q}$, $embd_{va}$, and $wgt_{q}$ denote the quadrant embedding, VA embedding, and the weight assigned to the quadrant, respectively. We set $weight_{q}$ to 0.5 and use cross-attention to fuse the emotion and music embeddings:
\begin{align}
EM = Attn(Q_{e},K_{m},V_{m}) = softmax(\frac{Q_{e}K_{m}^T}{\sqrt{d_k}})V_{m}
\end{align}
where $Q_{e}$, $K_{m}$, and $V_{m}$ represent the respective matrix for query (emotion embedding), key (music embedding), and value (music embedding), $d_{k}$ is the size of a key vector, and $EM$ is the emotion-aligned music embedding.

\begin{figure}[ht!]
  \centering
  \includegraphics[width=\columnwidth]{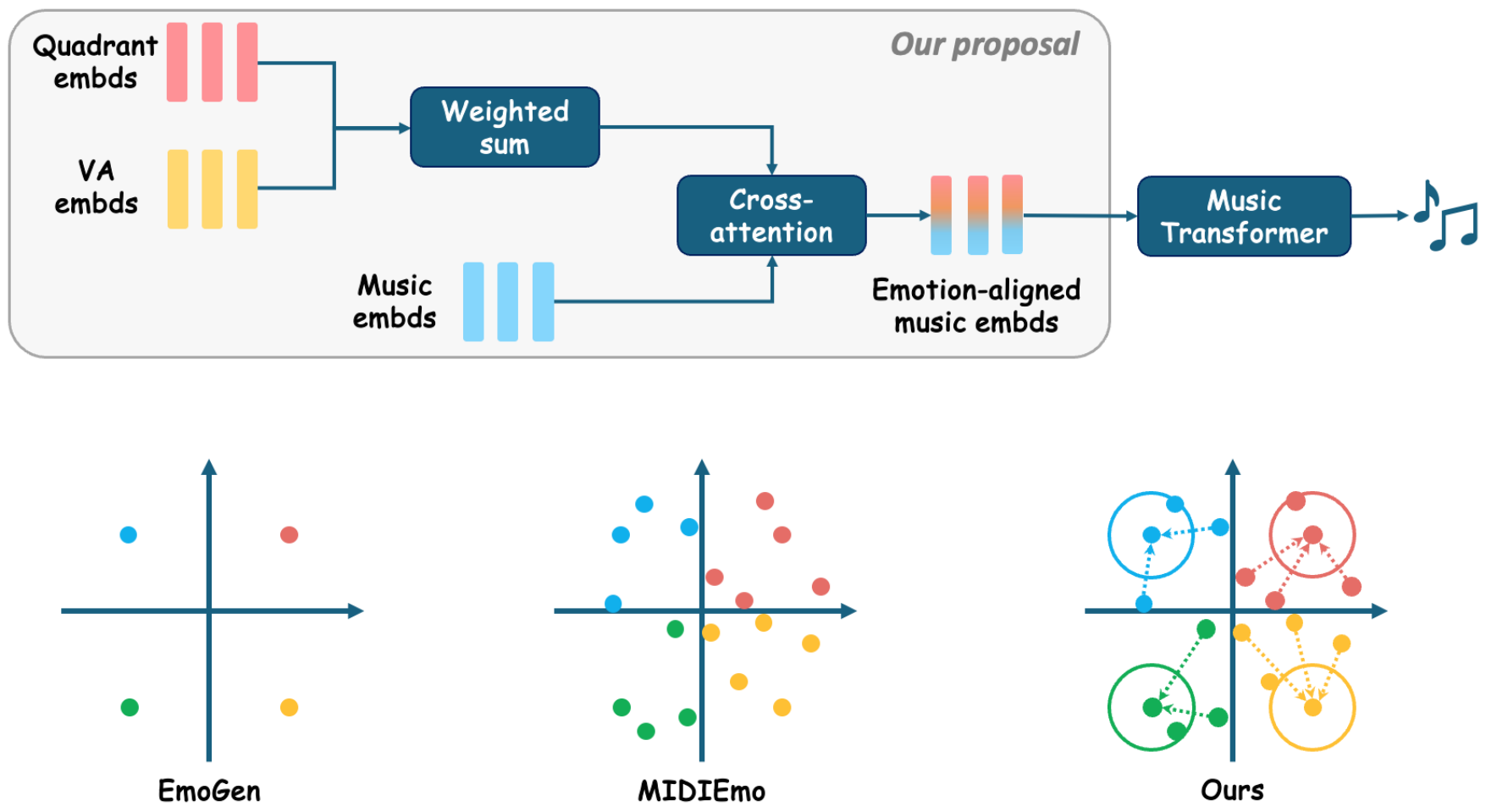}
  \caption{Our proposed EMG model (above) and its comparison to baseline models in terms of emotion conditioning (below).}
  \label{fig:models}
\end{figure}

Fig.~\ref{fig:models} illustrates our proposed model and compares it to baseline models in terms of emotion conditioning principles. It is evident that \textit{our approach} enhances emotion conditioning by combining the strengths of \textit{EmoGen} and \textit{MIDIEmo}, enabling a balance and trade-off between \textbf{prominence} and \textbf{variability} in the generated emotion.

The emotion is anchored at the quadrant center by $wgt_{q} \cdot embd_{q}$ and allows for variation within a certain range by $(1-wgt_{q}) \cdot embd_{va}$. Samples outside this range are adjusted to fall within, ensuring more pronounced emotion. The value of $wgt_{q}$ influences the size of this range: a larger $wgt_{q}$ results in a smaller range and emotion closer to the quadrant, while a smaller $wgt_{q}$ leads to a larger range, potentially reaching the quadrant boundary. By adjusting $wgt_{q}$, we provide flexibility for emotion to span between precision and variation.

For the remaining of our model, we follow \cite{sulun2022symbolic}, leveraging Music Transformer \cite{huang2018music} as the backbone and using its Lakh-Spotify dataset based on Lakh Pianoroll 5 full dataset \cite{dong2018musegan} for model training. For brevity, we omit the details and refer readers to Sec. III of \cite{sulun2022symbolic}.

\subsection{Comparative Evaluation -- Real vs. Synthetic Music}
To assess the quality of generated music, prior research has employed either subjective or objective evaluations. In subjective evaluations, human raters assign scores to the generated music based on the accuracy of the conveyed emotions. In objective evaluations, an emotion classifier is used to classify the generated music \cite{kang2023emogen,ferreira2022controlling}. However, both approaches are subject to inherent biases, either from human perception or emotion classifiers. Therefore, we propose an objective measurement based on FAD.

We generate music using \textit{MIDIEmo}, \textit{EmoGen}, and our proposed model, producing a total of 1,000 music samples from each model\footnote{Samples available: \href{https://yc-li20.github.io/Music-Emotion}{https://yc-li20.github.io/Music-Emotion}}. For \textit{EmoGen}, we generate 250 samples for each of the four emotion quadrants. For \textit{MIDIEmo} and our model, which encode continuous VA values, we generate 250 samples for VA pairs in each quadrant with a balanced distribution. Since all three models produce MIDI music, we use EMOPIA as the reference dataset. EMOPIA, also in MIDI format, comprises approximately 1,000 samples with a balanced distribution across the four quadrants. Consistent with Section~\ref{sec:fad}, we utilize all audio encoders and calculate their average FAD scores to mitigate the emotion bias. The comparison of FAD scores across different quadrants is presented in Table~\ref{fig:fadcomparison}. The observations are as follows:

\begin{table}[ht!]
\centering
\large
\caption{FAD comparison of Real and Synthetic Music.}
\scalebox{0.74}{
\begin{tabular}{lcccccc}
\hline
 & \textbf{\normalsize{Q1\_Q2}} & \textbf{\normalsize{Q1\_Q3}} & \textbf{\normalsize{Q1\_Q4}} & \textbf{\normalsize{Q2\_Q3}} & \textbf{\normalsize{Q2\_Q4}} & \textbf{\normalsize{Q3\_Q4}} \\ \hdashline
\textit{\normalsize{EMOPIA (real)}} & \large{1.36} & \large{11.60} & \large{11.24} & \large{13.64} & \large{13.18} & \large{1.47} \\
\textit{\normalsize{EmoGen (syn.)}} & \large{9.26} & \large{60.06} & \large{30.42} & \large{74.64} & \large{49.45} & \large{26.15} \\
\textit{\normalsize{MIDIEmo (syn.)}} & \large{1.67} & \large{51.68} & \large{35.62} & \large{51.42} & \large{34.93} & \large{3.70} \\ \hdashline
\textit{\normalsize{Ours (syn.)}} & \large{1.61} & \large{33.13} & \large{26.10} & \large{17.93} & \large{15.96} & \large{5.99} \\ \hline
\end{tabular}}
\label{fig:fadcomparison}
\end{table}

\textit{\textbf{1)}} The FAD scores for Q1\_Q2 and Q3\_Q4 are notably lower compared to other quadrant pairs. As discussed in Sec.~\ref{sec:fad}, this is likely because the emotions in these quadrant pairs share the same arousal polarity, making it more challenging to distinguish between them.

\textit{\textbf{2)}} \textit{EmoGen} achieves the highest FAD scores, indicating that it excels at differentiating between various emotions based on their musical characteristics. This is consistent with its training principle, where all music pieces within a given quadrant are conditioned to express the same emotion. However, this practice results in less emotional variation, making the generated emotions appear less natural. An advantage of \textit{EmoGen} is its effective differentiation for Q1\_Q2 and Q3\_Q4, as it achieves higher scores than those of real music.

\textit{\textbf{3)}} \textit{MIDIEmo} exhibits greater variability than \textit{EmoGen}. Although its overall FAD scores are lower than those of \textit{EmoGen}, they are still higher than those for real music, particularly for Q1\_Q3 and Q2\_Q4. Since emotions in diagonal quadrants are inherently more distinct, a high score in these cases might cause the generated emotional music to sound overly pronounced or exaggerated.

\textit{\textbf{4)}} Our overall FAD scores are closer to those of real music compared to \textit{EmoGen} and \textit{MIDIEmo}, indicating that our model generates more realistic emotions. Furthermore, the score for Q3\_Q4 in our model is higher than that of \textit{MIDIEmo} and real music, demonstrating that our approach effectively leverages the strengths of \textit{EmoGen}. As discussed in Sec.~\ref{sec:fad}, relatively high scores for Q1\_Q2 and Q3\_Q4 are beneficial for distinguishing between these quadrants. Additionally, the emotion embedding in our model is naturally aligned with the music embedding through cross-attention, resulting in overall lower scores compared to \textit{MIDIEmo}, which simply concatenates emotion and music embeddings.

\section{Conclusion}
In this study, we address the issue of emotion bias in music, focusing on both its recognition and generation. We benchmark MER across diverse music datasets, highlighting discrepancies between categorical and dimensional emotions, as well as the limitations of relying on a single encoder or metric description model. To address these issues, we propose employing FAD with various audio encoders as an objective measurement. Additionally, we introduce an enhanced model for emotion generation that integrates the advantages of both categorical and dimensional emotions, resulting in more realistic emotional expression in the generated music, with control over both prominence and variability. The comparative evaluation of real and synthetic music using FAD demonstrates the effectiveness of our approach. This study underscores the issue of emotion bias in music and provides novel insights into addressing this challenge.

\section*{Acknowledgment}
We thank Xu Tan (former Principal Research Manager), Peiling Lu (former Research Software Development Engineer), and Chenfei Kang (former intern) from Microsoft Research Asia for their assistance in implementing \textit{EmoGen}, and Serkan Sulun from INESC TEC for his help in implementing \textit{MIDIEmo}.

Yuanchao appreciates the feedback from all members of the Audio and Acoustics Research Group at Microsoft Research, as well as the great summer spent with fellow interns Ard Kastrati, Azalea Gui, Eloi Moliner Juanpere, Michele Mancusi, Ruihan Yang, and Tanmay Srivastava.

\balance
\bibliographystyle{IEEEbib}
\bibliography{icme2025references}

\begin{thebibliography}{10}

\bibitem{hung2021emopia}
Hsiao-Tzu Hung, Joann Ching, Seungheon Doh, Nabin Kim, Juhan Nam, and Yi-Hsuan Yang,
\newblock ``{EMOPIA}: A multi-modal pop piano dataset for emotion recognition and emotion-based music generation,''
\newblock in {\em ISMIR}, 2021.

\bibitem{yang2011music}
Yi-Hsuan Yang and Homer~H Chen,
\newblock {\em Music emotion recognition},
\newblock CRC Press, 2011.

\bibitem{hu2007exploring}
Xiao Hu and J~Stephen Downie,
\newblock ``Exploring mood metadata: Relationships with genre, artist and usage metadata.,''
\newblock in {\em ISMIR}, 2007.

\bibitem{sulun2022symbolic}
Serkan Sulun, Matthew~EP Davies, and Paula Viana,
\newblock ``Symbolic music generation conditioned on continuous-valued emotions,''
\newblock {\em IEEE Access}, vol. 10, pp. 44617--44626, 2022.

\bibitem{krumhansl2002music}
Carol~L Krumhansl,
\newblock ``Music: A link between cognition and emotion,''
\newblock {\em Current directions in psychological science}, vol. 11, no. 2, pp. 45--50, 2002.

\bibitem{hallam2009oxford}
Susan Hallam, Ian Cross, and Michael Thaut,
\newblock {\em Oxford handbook of music psychology},
\newblock Oxford University Press, 2009.

\bibitem{hunter2010music}
Patrick~G Hunter and E~Glenn Schellenberg,
\newblock ``Music and emotion,''
\newblock {\em Music perception}, pp. 129--164, 2010.

\bibitem{chen2018cross}
Yi-Wei Chen, Yi-Hsuan Yang, and Homer~H Chen,
\newblock ``Cross-cultural music emotion recognition by adversarial discriminative domain adaptation,''
\newblock in {\em 2018 17th IEEE International Conference on Machine Learning and Applications (ICMLA)}. IEEE, 2018, pp. 467--472.

\bibitem{russell1980circumplex}
James~A Russell,
\newblock ``A circumplex model of affect.,''
\newblock {\em Journal of personality and social psychology}, 1980.

\bibitem{kang2023emogen}
Chenfei Kang, Peiling Lu, Botao Yu, Xu~Tan, Wei Ye, Shikun Zhang, and Jiang Bian,
\newblock ``{EmoGen}: Eliminating subjective bias in emotional music generation,''
\newblock {\em arXiv preprint arXiv:2307.01229}, 2023.

\bibitem{song2016perceived}
Yading Song, Simon Dixon, Marcus~T Pearce, and Andrea~R Halpern,
\newblock ``Perceived and induced emotion responses to popular music: Categorical and dimensional models,''
\newblock {\em Music Perception: An Interdisciplinary Journal}, vol. 33, no. 4, pp. 472--492, 2016.

\bibitem{madhok2018sentimozart}
Rishi Madhok, Shivali Goel, and Shweta Garg,
\newblock ``{SentiMozart}: Music generation based on emotions.,''
\newblock in {\em ICAART (2)}, 2018, pp. 501--506.

\bibitem{soleymani20131000}
Mohammad Soleymani, Micheal~N Caro, Erik~M Schmidt, Cheng-Ya Sha, and Yi-Hsuan Yang,
\newblock ``1000 songs for emotional analysis of music,''
\newblock in {\em 2nd ACM international workshop on Crowdsourcing for multimedia}, 2013.

\bibitem{panda2013multi}
Renato Eduardo~Silva Panda, Ricardo Malheiro, Bruno Rocha, Ant{\'o}nio~Pedro Oliveira, and Rui~Pedro Paiva,
\newblock ``Multi-modal music emotion recognition: A new dataset, methodology and comparative analysis,''
\newblock in {\em CMMR}, 2013.

\bibitem{downie20082007}
John~Stephen Downie, Cyril Laurier, and Andreas Ehmann,
\newblock ``The 2007 mirex audio mood classification task: Lessons learned,''
\newblock in {\em Proc. 9th Int. Conf. Music Inf. Retrieval}, 2008, pp. 462--467.

\bibitem{hershey2017cnn}
Shawn Hershey, Sourish Chaudhuri, Daniel~PW Ellis, Jort~F Gemmeke, Aren Jansen, R~Channing Moore, Manoj Plakal, Devin Platt, Rif~A Saurous, Bryan Seybold, et~al.,
\newblock ``Cnn architectures for large-scale audio classification,''
\newblock in {\em 2017 ieee international conference on acoustics, speech and signal processing (icassp)}. IEEE, 2017, pp. 131--135.

\bibitem{elizalde2024natural}
Benjamin Elizalde, Soham Deshmukh, and Huaming Wang,
\newblock ``Natural language supervision for general-purpose audio representations,''
\newblock in {\em ICASSP}. IEEE, 2024.

\bibitem{laionclap2023}
Yusong Wu, Ke~Chen, Tianyu Zhang, Yuchen Hui, Taylor Berg-Kirkpatrick, and Shlomo Dubnov,
\newblock ``Large-scale contrastive language-audio pretraining with feature fusion and keyword-to-caption augmentation,''
\newblock in {\em ICASSP}. IEEE, 2023.

\bibitem{li2023mert}
Yizhi Li, Ruibin Yuan, Ge~Zhang, Yinghao Ma, Xingran Chen, Hanzhi Yin, Chenghao Xiao, Chenghua Lin, Anton Ragni, Emmanouil Benetos, et~al.,
\newblock ``{MERT}: Acoustic music understanding model with large-scale self-supervised training,''
\newblock in {\em ICLR}, 2023.

\bibitem{Manocha2021CCL}
Pranay Manocha, Zeyu Jin, Richard Zhang, and Adam Finkelstein,
\newblock ``{CDPAM}: Contrastive learning for perceptual audio similarity,''
\newblock in {\em ICASSP}. IEEE, 2021.

\bibitem{defossez2022highfi}
Alexandre D{\'e}fossez, Jade Copet, Gabriel Synnaeve, and Yossi Adi,
\newblock ``High fidelity neural audio compression,''
\newblock {\em Transactions on Machine Learning Research}, 2023.

\bibitem{kumar2024high}
Rithesh Kumar, Prem Seetharaman, Alejandro Luebs, Ishaan Kumar, and Kundan Kumar,
\newblock ``High-fidelity audio compression with improved rvqgan,''
\newblock {\em Advances in Neural Information Processing Systems}, 2024.

\bibitem{castellon2021codified}
Rodrigo Castellon, Chris Donahue, and Percy Liang,
\newblock ``Codified audio language modeling learns useful representations for music information retrieval,''
\newblock in {\em ISMIR}, 2021.

\bibitem{kim2010music}
Youngmoo~E Kim, Erik~M Schmidt, Raymond Migneco, Brandon~G Morton, Patrick Richardson, Jeffrey Scott, Jacquelin~A Speck, and Douglas Turnbull,
\newblock ``Music emotion recognition: A state of the art review,''
\newblock in {\em ISMIR}, 2010.

\bibitem{droit2013music}
Sylvie Droit-Volet, Danilo Ramos, Jos{\'e}~LO Bueno, and Emmanuel Bigand,
\newblock ``Music, emotion, and time perception: the influence of subjective emotional valence and arousal?,''
\newblock {\em Frontiers in Psychology}, 2013.

\bibitem{li2019expressing}
Yuanchao Li, Carlos~Toshinori Ishi, Koji Inoue, Shizuka Nakamura, and Tatsuya Kawahara,
\newblock ``Expressing reactive emotion based on multimodal emotion recognition for natural conversation in human--robot interaction,''
\newblock {\em Advanced Robotics}, vol. 33, no. 20, pp. 1030--1041, 2019.

\bibitem{mcpherson2018role}
Laura McPherson,
\newblock ``The role of music in documenting phonological grammar: Two case studies from west africa,''
\newblock in {\em Proceedings of the Annual Meetings on Phonology}, 2018.

\bibitem{li2023exploration}
Yuanchao Li, Yumnah Mohamied, Peter Bell, and Catherine Lai,
\newblock ``Exploration of a self-supervised speech model: A study on emotional corpora,''
\newblock in {\em SLT}. IEEE, 2023.

\bibitem{saliba2024layer}
Alexandra Saliba, Yuanchao Li, Ramon Sanabria, and Catherine Lai,
\newblock ``Layer-wise analysis of self-supervised acoustic word embeddings: A study on speech emotion recognition,''
\newblock in {\em ICASSP SASB Workshop}. IEEE, 2024.

\bibitem{kilgour2019frechet}
Kevin Kilgour, Mauricio Zuluaga, Dominik Roblek, and Matthew Sharifi,
\newblock ``Fr{\'e}chet audio distance: A reference-free metric for evaluating music enhancement algorithms.,''
\newblock in {\em INTERSPEECH}, 2019, pp. 2350--2354.

\bibitem{gui2023adapting}
Azalea Gui, Hannes Gamper, Sebastian Braun, and Dimitra Emmanouilidou,
\newblock ``Adapting frechet audio distance for generative music evaluation,''
\newblock {\em ICASSP}, 2024.

\bibitem{sun2025exploring}
Yujia Sun, Zeyu Zhao, Korin Richmond, and Yuanchao Li,
\newblock ``Exploring acoustic similarity in emotional speech and music via self-supervised representations,''
\newblock in {\em ICASSP 2025-2025 IEEE International Conference on Acoustics, Speech and Signal Processing (ICASSP)}. IEEE, 2025, pp. 1--5.

\bibitem{hong2017analysis}
Yu~Hong, Chuck-Jee Chau, and Andrew Horner,
\newblock ``An analysis of low-arousal piano music ratings to uncover what makes calm and sad music so difficult to distinguish in music emotion recognition,''
\newblock {\em Journal of the Audio Engineering Society}, 2017.

\bibitem{huang2018music}
Cheng-Zhi~Anna Huang, Ashish Vaswani, Jakob Uszkoreit, Ian Simon, Curtis Hawthorne, Noam Shazeer, Andrew~M Dai, Matthew~D Hoffman, Monica Dinculescu, and Douglas Eck,
\newblock ``Music transformer: Generating music with long-term structure,''
\newblock in {\em International Conference on Learning Representations}, 2018.

\bibitem{dong2018musegan}
Hao-Wen Dong, Wen-Yi Hsiao, Li-Chia Yang, and Yi-Hsuan Yang,
\newblock ``Musegan: Multi-track sequential generative adversarial networks for symbolic music generation and accompaniment,''
\newblock in {\em AAAI}, 2018.

\bibitem{ferreira2022controlling}
Lucas~N Ferreira, Lili Mou, Jim Whitehead, and Levi~HS Lelis,
\newblock ``Controlling perceived emotion in symbolic music generation with monte carlo tree search,''
\newblock in {\em AAAI}, 2022.

\end{thebibliography}

\end{document}